# Origin of high piezoelectricity at the morphotropic phase boundary (MPB) in $(Pb_{0.94}Sr_{0.06})(Zr_xTi_{1-x})O_3$

Ravindra Singh Solanki,[1§] Sunil Kumar Mishra,[2] Chikako Moriyoshi,[3] Yoshihiro Kuroiwa,[3] Isao Ishii, [4] Takashi Suzuki,[4] and Dhananjai Pandey[1*]

[1]School of Materials Science and Technology, Indian Institute of Technology (Banaras HinduUniversity), Varanasi-221005, India

[2]UGC-Academic Staff College, Dr. H. S. Gour Vishwavidyalaya, Sagar-470003, Madhya Pradesh, India

[3]Department of Physical Science, Graduate School of Science, Hiroshima University, Japan

[4]Department of Quantum Matter, ADSM, Hiroshima University, Japan

## Abstract

In this work, we address the issue of peaking of piezoelectric response at a particular composition in the morphotropic phase boundary (MPB) region of $(Pb_{0.940}Sr_{0.06})(Zr_xTi_{1-x})O_3$ (PSZT) piezoelectric ceramics. We present results of synchrotron x-ray powder diffraction, dielectric, piezoelectric and sound velocity studies to critically examine the applicability of various models for the peaking of physical properties. It is shown that the models based on the concepts of phase coexistence, polarization rotation due to monoclinic structure, tricritical point and temperature dependent softening of elastic modulus may enhance the piezoelectric response in the MPB region in general but cannot explain its peaking at a specific composition. Our results reveal that the high value of piezoelectric response for the MPB compositions in PSZT at x=0.530 is due to the softening of the elastic modulus as a function of composition. The softening of elastic modulus facilitates the generation of large piezoelectric strain and polarization on approaching the MPB composition of x=0.530. This new finding based on the softening of elastic modulus may pave the way forward for discovering/designing new lead-free environmentally friendly piezoelectric materials and revolutionize the field of piezoelectric ceramics.

***Corresponding Author's email:** dp.mst1979@gmail.com

§**Present affiliation:** Centre of Material Sciences, Institute of Interdisciplinary Studies, University of Allahabad, Allahabad-211002, Uttar Pradesh, India

## I. Introduction

Lead-based piezoelectric ceramics, such as $Pb(Zr_xTi_{1-x})O_3$ (PZT), $Pb(Mg_{1/3}Nb_{2/3})_xTi_{1-x}O_3$ (PMN-xPT) and $Pb(Zn_{1/3}Nb_{2/3})_xTi_{1-x}O_3$ (PZN-xPT), are known to exhibit ultrahigh piezoelectric response for compositions close to a first order phase boundary (MPB) across which a composition induced structural phase transition, commonly known as morphotropic phase transition, accompanied with a phase coexistence region occurs at room temperature. Morphotropic phase boundary (MPB) compositions of PZT and other two ceramics form the backbone of most of the present day electromechanical devices [1] as the piezoelectric response shows a peak corresponding to an MPB composition. Several models based on the coexistence of phases in the MPB region [2], lattice instability near room temperature [3, 4], existence of a tricritical point near MPB in PZT [5, 6] giving rise to flat energy surfaces [7, 8] for the rotation of the polarization vector [9-14] have been proposed to explain the origin of ultrahigh piezoelectric response in the MPB region. However, as evident from recent publications [7, 8], the issue of high piezoelectric response at MPB is an open issue. The discovery of the monoclinic phases of $M_A$ (space group Cm [15, 16]), $M_B$ (space group Cm [17]) and $M_C$ (space group Pm [17]) types in the MPB region of lead-based piezoelectric ceramics, such as PZT, PMN-xPT and PZN-xPT, has brought in another view point for the ultrahigh piezoelectric response at the MPB as these phases permit the rotation of the polarization vector on a symmetry plane in response to an external electric field [10, 11] unlike the orthorhombic, tetragonal and rhombohedral phases where the polarization vector lies along fixed [110], [001] and [111] directions, respectively. There is a recent surge of interest in the field of piezoelectric ceramics to discover environmentally friendly new lead (Pb)-free MPB systems with high piezoelectric response comparable to or better than that of the PZT based systems [18]. It is anticipated that a proper understanding of the origin of high piezoelectric response in PZT based ceramics may be helpful in designing new environmentally friendly piezoelectric ceramics with MPB characteristics [7, 18].

The present work was undertaken to critically evaluate the various models proposed for the ultrahigh piezoelectric response at the MPB in PZT based ceramics. We have selected 6% $Sr^{2+}$ modified PZT, i.e. PSZT, a PZT based ceramic whose ground state, high temperature phase stabilities and phase transitions have been investigated recently in great detail [6, 19] and provide a strong data base to test the various models of high piezoelectricity proposed so far in the context of pure PZT ceramics. We have examined the issue of high piezoelectric response at the MPB of PSZT through a study of (i) the structure of the crystallographic phases at room temperature across the MPB using high resolution synchrotron powder XRD, (ii) the physical properties such as relative permittivity ($\varepsilon'$), planar electromechanical coupling coefficient ($k_p$), piezoelectric strain coefficient ($d_{33}$) and longitudinal elastic modulus ($C_L$) at room temperature as a function of composition across the MPB and (iii) the temperature dependence of longitudinal elastic modulus ($C_L$) across the MPB. Our results show that the high value of piezoelectric response for x=0.530 in the MPB region is primarily linked with the extreme softening of the elastic modulus as a function of the composition at room temperature that drives the morphotropic phase transition. All other factors, like phase coexistence, flat energy surface for polarization rotation due to tricritical point, existence of monoclinic phases and temperature dependent lattice instability in the vicinity of the PZT, may have a secondary role, if at all, in causing a general increase in the piezoelectric activity near the MPB compositions but not the peak in the piezoelectric response for a specific composition. This finding may provide an insight for designing new lead-free piezoelectric ceramics that are environmentally friendly in contrast to the toxic Pb-based MPB ceramics.

**II. Experimental**

Single phase powders of $(Pb_{0.94}Sr_{0.06})(Zr_xTi_{1-x})O_3$ (henceforth abbreviated as PSZT) were obtained by solid-state thermochemical reaction in a stoichiometric mixture of $(Pb_{0.94}Sr_{0.06})CO_3$ and $(Zr_xTi_{1-x})O_2$ powder for $x$ = 0.515, 0.520, 0.525, 0.530, 0.535, 0.545, and 0.550 at $800^0$C for six hours where the carbonate and zirconia solid solutions were obtained by chemical routes that gives very homogeneous distribution of $Zr^{4+}$ and $Ti^{4+}$and $Pb^{2+}$ and $Sr^{2+}$ cations at the "A" and "B" sites of the $ABO_3$ perovskite structure, respectively. This semi-wet method of synthesizing PZT ceramics gives the narrowest MPB region with $\Delta x \sim 0.01$ [20, 21] on account of very good chemical homogeneity at the "B" site in contrast to all solid state route or other wet-routes [22, 23]. Full description of sample preparation technique for PSZT has been given in our previous works [19]. The microstructure of the sintered ceramics was studied using a field emission gun based scanning electron microscope (SEM) (Supra 40, Zeiss, Germany). The average grain size of the sintered pellets using linear intercept method is found to be ~4.3μm (see section S.1 of the supplementary file).

Fired-on silver electrodes were applied after polishing the surfaces of the PZT pellets with 0.25μm diamond paste. Poling was performed by applying dc field of approximately 20kV/cm for 50 minutes at $100^0$C in a silicon oil bath. The sample was then slowly cooled to room temperature with the dc field applied. The poled pellets were aged for at least 24 hours at room temperature before performing the electrical measurements. The room temperature values of dielectric constant on unpoled samples, planar electromechanical coupling coefficient ($k_p$) and $d_{33}$ of poled pellets were determined to locate the MPB. The dielectric and resonance-antiresonance frequency measurements were performed using a Schlumberger SI-1260 impedance gain phase analyser. The planar coupling coefficient $k_p$ was determined from the resonance and antiresonance frequencies as per the IRE specifications [23]. The piezoelectric strain coefficient ($d_{33}$) at room temperature was measured using a Berlincourt $d_{33}$ piezometer (Model CADT) at an operating frequency of 25Hz. The longitudinal elastic modulus ($C_L$) was obtained by measuring the sound velocity ($v$) using phase comparison type pulse echo method [24]. The elastic modulus $C$ was calculated using $C = \rho v^2$ relation, where $\rho$ is the room-temperature mass density of the sintered sample.

High-resolution synchrotron x-ray powder diffraction (SXRPD) measurements were carried out at room temperature at BL02B2 beam line of SPring-8, Japan at a wavelength of 0.412 Å (30 keV) [25]. Rietveld refinements were carried out using FULLPROF software package [26].

## III. Results and Discussion:

### A. Location of the MPB

We first locate the MPB region of PSZT at room temperature by studying the change in crystal structure of PSZT as a function of composition (x) at room temperature through a qualitative interpretation of the SXRPD data for the seven PSZT compositions under investigation in this work. We show the evolution of the $(111)_{pc}$, $(200)_{pc}$ and $(220)_{pc}$ (here pc stands for pseudocubic) perovskite reflections with composition in Fig.1. It is evident from these profiles that for x= 0.515, the $(111)_{pc}$ reflection is a singlet, while $(200)_{pc}$ and $(220)_{pc}$ are doublets. This is the characteristics of a tetragonal phase. On increasing the Zr-content from $x$ = 0.515, the width of the $(111)_{pc}$ peak starts increasing until, for x= 0.530, this peak splits and becomes an apparent doublet, while $(200)_{pc}$ is still a doublet. $(220)_{pc}$ peak clearly splits into three for compositions with x=0.530 and 0.535. These are the characteristics of the pseudotetragonal monoclinic phase in the Cm space group for the MPB compositions of PZT and PSZT [21, 27, 28]. With further increase in Zr-content, the $(200)_{pc}$ peak splitting starts disappearing and it becomes nearly a singlet but with a full width at half maximum (FWHM) that is ~2.66 times greater than that of the individual $(111)_{pc}$ peaks for x=0.550. Further, the $(111)_{pc}$ and $(220)_{pc}$ reflections become doublets. These are the characteristics of the 'pseudorhombohedral monoclinic phase' in the Cm space group discovered by Ragini et al. [27] and Singh et al. [28] in pure PZT and confirmed in PSZT also [19]. Thus, there is a morphotropic phase transition from the tetragonal structure stable for x≤0.515 to the pseudorhombohedral (PR) monoclinic structure stable for x≥0.545 through a pseudotetragonal (PT) monoclinic structure in the composition range 0.515<x<0.545.

Fig. 2 plots the room temperature values of dielectric constant, electromechanical coupling coefficient ($k_p$), piezoelectric strain constant ($d_{33}$), longitudinal ($C_L$) elastic modulus of PSZT as a function of composition (x). For the sake of completeness, we have included results of Mishra [29] on pure PZT in Fig. 2. As $Sr^{2+}$ substitution decreases the Curie temperature by $9.5^0$C per mole % of atom added, this leads to an increase in the dielectric constant of PSZT at room temperature in comparison to PZT [1]. Further, the increase in dielectric constant raises the $d_{33}$ for PSZT as compared to PZT, as per the relationship $d \sim 2 P_s \chi Q$ where $P_s$, $\chi$ and Q are the spontaneous polarization, dielectric susceptibility and electrostrictive coefficient [30]. The electromechanical coupling factor ($k_p$), on the other hand, is not greatly enhanced as a result of $Sr^{2+}$ substation but is found to peak at $x \simeq 0.530$ as compared to x~0.520 in PZT as shown in Fig. 3 (a), (b) and (c). In case of pure PZT ceramics, they peak at $x \approx 0.520$ [1, 29]. Our values of dielectric constant, electromechanical coupling coefficient and piezoelectric strain constant are in close to those reported by Lal et al [31]. Both the structural and physical property results indicate that the peaking of the piezoelectric response occurs at x=0.530 in PSZT and is linked with the morphotropic phase transition that occurs across x~0.530. We now proceed to evaluate the various models of high piezoelectricity at x~ 0.530 in the MPB composition region of PSZT.

**B. Role of phase coexistence on the piezoelectric response**

Isupov [2] had proposed that the coexistence of tetragonal and rhombohedral phases in the MPB region of PZT provides a large number of ferroelectric domain orientations (6 for tetragonal and 8 for rhombohedral phase), some of which can always respond more easily to external electric field than in pure tetragonal and rhombohedral phases with fewer domain orientations. According to Isupov [2], this may be responsible for the high piezoelectric response at the MPB. This proposition of Isupov was based on the belief in the early literature that there is a tetragonal to rhombohedral morphotropic phase transition across the MPB of the PZT ceramics [1, 5] and the width of this MPB region ($\Delta x$) was reported to be as wide as $\approx 0.15$ [22] but with significant improvements in the synthesis techniques it's intrinsic width was reported to be around 0.01 [5, 20]. After the recent work of Noheda et al [15, 16], who showed the presence of a monoclinic phase (Cm space group) in the MPB region, and Ragini et al [27] and Singh et al, [28] who showed that even the so-called rhombohedral compositions are monoclinic, it is now well established that the MPB region in PZT separates the tetragonal and monoclinic phases. However, there is still a coexistence of these two phases and therefore the possibility of many more domain orientations. We therefore carried out Rietveld refinements for various PSZT compositions showing single phase and two phase structures to determine the intrinsic width of the two phase region on account of the first order character of the morphotropic phase transition and its role on the high piezoelectricity at the MPB. The details of refinements are presented in section S.2 of the supplementary file and in the following section we discuss relevant results only.

In order to critically evaluate the applicability of the phase coexistence model, we show in Fig. 3 the variation of phase fractions and the pseudocubic lattice parameters ($a_p$, $b_p$, $c_p$) with composition across the MPB region, as determined by Rietveld refinements. For the MPB composition range, we have used subscript '1' for the 'pseudorhombohedral' monoclinic phase and '2' for the tetragonal phase for $x \leq 0.525$ and to the 'pseudotetragonal' monoclinic phase for $x > 0.525$. Variation of the pseudocubic lattice parameters of pseudotetragonal and pseudorhombohedral monoclinic phases show a similar dependence on composition as reported earlier by Singh et al. [28] for pure

PZT. The pseudocubic $c_{p1}/a_{p1}$ ratio (see Fig. 3(c)) for pseudorhombohedral phase decreases sharply with increasing Zr-content and becomes close to 1 for x≥0.545. This is consistent with the results of Rietveld refinement that the nature of the monoclinic phase changes from pseudotetragonalfor 0.520≤x<0.545 to pseudorhombohedral for x≥0.545. Pseudotetragonality of phase '2' also decreases on increasing Zr-content. Pseudocubic volumes for phase '1' ($V_{p1}$) and '2' ($V_{p2}$) increase with increasing value of Zr-content and saturate for x>0.535 and x>0.525, respectively. It is evident from Fig. 3(a) that the tetragonal phase first transforms to the pseudotetragonal monoclinic phase to minimize the strains at the interface and it coexists with the pseudotetragonal monoclinic phase for 0.515 < x < 0.530. With increasing $Zr^{4+}$ content, the phase fraction of the tetragonal phase decreases and becomes zero for x=0.530 while the phase fraction of thepseudotetragonal monoclinic phase increases and becomes maximum for x=0.530. On increasing the $Zr^{4+}$ content further, the phase fraction of the pseudotetragonal monoclinic phase decreases and becomes zero for x≥0.545. This phase coexists with the pseudorhombohedral monoclinic phase for 0.530 < x< 0.545. For x≥0.545, the structure corresponds to pure pseudorhombohedral phase. Thus, the pseudotetragonal monoclinic phase gives way to pure pseudorhombohedral monoclinic phase with increasing $Zr^{4+}$ content through a narrow composition region of coexistence of the two monoclinic (ie, pseudotetragonal and pseudorhombohedral) phases to minimize the strains at the interface of the two phases, as can be seen from the gradual decrease in $c_{p2}/a_{p2}$ ratio of the pseudotetragonal monoclinic phase from 1.0145 for x=0.520 to 1.0047 for x=0.535.

It is evident from the above discussion that the tetragonal and pseudotetragonal monoclinic phases coexist for 0.515 < x < 0.530 while pseudotetragonal and pseudorhombohedral monoclinic phases coexist for 0.530 > x > 0.545. The phase coexistence region in PSZT extends from x≈0.520 to x≈0.535 whereas the peak in the physical properties occurs at a specific composition for x=0.530.In this context, it is interesting to note that the fraction of the pseudotetragonal monoclinic phase is maximum for x=0.530 for which all the properties also show a peak. However, this is not true in PZT where the pseudotetragonal monoclinic phase fraction peaks at 0.525 whereas the peak in properties occurs at x=0.520. This suggests that the coexistence of two phases may not be the primary reason behind the peaking of properties at a specific composition in the MPB region. It probably involves deeper physics.

**C. Role of polarization rotation and flat energy surfaces in monoclinic phase:**

In the monoclinic structure, the polarization vector can lie anywhere on a symmetry plane [15, 16]. This may allow unrestricted rotation of the polarization vector from [001] of the tetragonal towards the [111] of the pseudorhombohedral phase through the monoclinic symmetry in a plane of symmetry of the Cm space group in response to an applied external electric field. Such a rotation of the polarization vector has been predicted theoretically [9-12] and confirmed experimentally [13]. It has been proposed that this characteristic of the monoclinic phase(s), present in the MPB region, may be responsible for the maximum piezoelectric response at the MPB. First principles calculations on PZT have indeed linked the high piezoelectric coefficient at the MPB due to the presence of the monoclinic phase that corresponds to what we call as the pseudotetragonal monoclinic phase [10, 11]. However, in PZT, as mentioned in the previous section, the maximum piezoelectric response occurs for a composition with x=0.520 which is mostly a tetragonal composition with a minority pseudotetragonal monoclinic phase coexisting with it [4]. Further, using a temperature dependent study of the monoclinic to tetragonal phase transition in the MPB region, it was shown by Singh et al. in PZT that the piezoelectric response of the monoclinic phase was in general less than that of the tetragonal phase for any composition in the MPB region [4]. Further, our present results on PSZT and previous results by Ragini et al [27] and Singh et al [28] on PZT reveal that the pseudotetragonal monoclinic phase is present in the entire MPB composition range (e.g.$0.520 \leq x \leq 0.535$ and $0.520 \leq x \leq 0.530$ for PSZT and PZT, respectively) in coexistence with either the tetragonal phase or a pseudorhombohedral monoclinic phase whereas the peak occurs at a specific composition x=0.530 for PSZT and x=0.520 for PZT. All these observations suggest that polarization rotation may not be the deciding factor in causing the peak in piezoelectric properties at a particular composition in the MPB region even though it may contribute to the general enhancement of the piezoelectric response in the MPB region where the monoclinic phases coexist.

**D. Role of flat energy surfaces for easy polarization rotation**

For polarisation rotation to occur in the monoclinic phase, the energy surface should be flat, as was shown theoretically by Cohen [9] and elaborated further by Damjanovic [7]. One of the ways for achieving such a flat energy surface is through the existence of a tricritical point at the MPB. This has been recently proposed to be the possible reason for the enhancement of piezoelectric activity in a lead free $BaTiO_3$ based pseudo binary MPB system $Ba(Zr_{0.2}Ti_{0.8})O_3$-$x(Ba_{.7}Ca_{.3})TiO3$ with a room temperature $d_{33}$ value of ~ 620 pC/N for the MPB composition at $x$=0.50 [32]. In this system, a tricritical point coinciding with a triple point at the terminal point of the MPB, where it meets the line of Curie transition, between the tetragonal and rhombohedral states has been reported and correlated with the highest piezoelectric response. Strongly piezoelectric lead-based systems such as PZT, PMN-xPT, PZN-xPT are also known to possess triple points in their phase diagrams [33] but the existence of a tricritical point coinciding with the triple point where the MPB intersects the ferroelectric to paraelectric line (Curie transition line) has been confirmed unambiguously only in PZT and PSZT ceramics for a composition corresponding to x=0.550 [5, 6]. Further, In the context of the PZT ceramics, it has been proposed that the high piezoelectric response of PZT is due to both the MPB effect which gives rise to polarization rotation due to presence of the monoclinic Cm phase and the proximity of the MPB with the tricritical point (coinciding with the triple point) which leads to flatter energy surface [7] Flatter energy surface indicates a higher susceptibility of the system to atomic displacements leading to the enhancement in dielectric, electromechanical and piezoelectric responses as was first shown by Fu and Cohen [9] using first principle calculations. Since the tricritical point in PZT based compositions lies well above the room temperature ($T_c \simeq 647$ and $\simeq 605$K in PZT [21] and PSZT [6, 19], respectively), the absolute flattening of the free energy surface on lowering the temperature to room temperature would disappear in these ceramics, in marked contrast to the $BaTiO_3$ based pseudo-binary system where $T_c$ is only marginally above the room temperature. Also, the compositions x=0.520 and 0.530 at which the piezoelectric properties peak at room temperature in PZT and PSZT ceramics, respectively, are away from the tricritical triple point composition (x=0.550) on account of the slightly tilted nature of the MPB towards the $Zr^{4+}$ rich side. It therefore remains questionable as to what extent the energy flattening would occur at room temperature for the MPB composition due to the tricritical point at the triple point. Some workers [8] using Landau theory considerations for PZT have proposed the presence of two tricritical points around x=0.30 and 0.80 that leaves paraelectric to ferroelectric phase transition second order for compositions lying in between these two tricritical points. A second-order phase transition can give rise to larger response functions in comparison to a first-order phase transition due to the flattening of the free-energy profile thus giving rise to high piezoelectric response of PZT in the MPB region. However, from this model, and the above discussion of the tricritical point based model, it is not evident why the piezoelectric response should peak only at a particular MPB composition and not for all the compositions between the two tricritical point compositions. So the role of $2^{nd}$ order Curie transition around MPB compositions or existence of a tricritical point at the triple point may not be the dominant factor in deciding the high piezoelectric response at x=0.520 and 0.530 in PZT and PSZT, respectively.

**E. Role of temperature dependent elastic instabilities on the piezoelectric response**

Soon after the discovery of the monoclinic phase in the Cm space group below the room temperature as a result of a phase transition from the tetragonal phase for x=0.520 by Noheda et al [34], Ragini et al [35] and Ranjan et al [36] showed the existence of another monoclinic phase resulting from the Cm phase due to an antiferrodistortive (AFD) phase transition through their electron and neutron diffraction measurements. The space group of this phase was subsequently confirmed as Cc [37-39]. Both the tetragonal to monoclinic (P4mm to Cm) and the AFD (Cm to Cc) transitions were shown to be preceded by the softening of the elastic modulus which hardens below the two transition temperatures ($T_{P4mm\text{-}Cm}$ and $T_{AFD}$) [35]. Elastic softening at the tetragonal to monoclinic

Cm and monoclinic Cm to monoclinic Cc transitions reported by Ragini et al [35] has been confirmed in a subsequent study by Cordero et al [40] using anelastic measurements as a function of temperature. It was noted by Singh et al [4] that only the elastic modulus of the tetragonal compositions showed softening around room temperature due to tetragonal to monoclinic Cm phase transition as pseudorhombohedral monoclinic compositions ( $x \geq 0.530$) underwent Cm to Cc transition well below room temperature. Taking a cue from this observation, Singh et al [4] proposed that the tetragonal to monoclinic transition temperature ($T_{P4mm\text{-}Cm}$) being closest to the room temperature for x= 0.520 in PZT with x=0.520 may be responsible for the high piezoelectric response at this composition as the softening of the modulus can cause large piezoelectric strain on application of an external field. We therefore critically examine the role of elastic instabilities as a function of temperature on the piezoelectric response of PSZT in this section.

Singh et al [4] derived the elastic modulus values of PZT from the piezoelectric resonance frequencies measured using poled ceramic samples. It is well known that ultrasonic measurement is a very sensitive probe for all kinds of phase transitions, including ferroelectric, magnetic and structural ones. We therefore carried out ultrasonic sound velocity measurements on PSZT samples by phase comparison type pulse echo method for the determination of the elastic modulus. In this method, pulsed ultrasound waves travel in the sample and are reflected back and forth within the sample [41]. This measurement has enabled us to measure the elastic modulus (C) which is given by $C = \rho v^2$, where ρ is room-temperature mass density (ρ) of the sintered sample and v the velocity of sound.

Fig. 4 depicts the variation of longitudinal elastic modulus ($C_L$) with temperature in the 4-300K temperature range for compositions with x=0.515, 0.520, 0.525, and 0.545. In our previous work for compositions with x=0.530 and 0.550, we showed that the anomaly in $C_L$ below the room temperature coincides with the AFD phase transition temperature between two ferroelectric monoclinic phases in the Cm and Cc space groups using neutron diffraction measurements [19]. It is evident from Fig. 2 (e) and 4 that the antiferrodistortive phase transition temperature ($T_{AFD}$) increases with the increasing Zr-content and shifts to the room temperature side. For easy correlation with the MPB effect, we have plotted both the tetragonal to monoclinic ($T_{T\text{-}M}$) phase transition temperature and $T_{AFD}$ near the MPB compositions of pure PZT and PSZT, respectively, as a function of composition (x) in Fig 1. It is evident that the $T_{AFD}$ of PSZT is always higher than PZT except for x=0.515.This is expected as $Sr^{2+}$ substitution reduces the average "A" site cationic radius and thereby promotes the rotational instability for AFD transition [42-44]. It is to evident from Fig. 1 that the $T_{T\text{-}M}$ lies around room temperature but $T_{AFD}$ lies below room temperature for the MPB compositions of PZT. However, according to the temperature dependence of the elastic modulus, the $T_{AFD}$ for Cm to Cc phase transition for PSZT is closest to room temperature for x=0.550 and shall approach room temperature (300K) for $x \simeq 0.573$. According to the model of Singh et al [4], the maximum value of piezoelectric response should therefore occur around $0.550 < x \lesssim 0.573$ in PSZT. However, as evident from Fig. 1, the peak occurs at x=0.530. Thus, the presence of elastic instability near room temperature may not be the primary factor for the peaking of the piezoelectric response in the PSZT.

**F. Role of elastic instability at room temperature as a function of composition:**

The composition dependence of longitudinal elastic modulus ($C_L$) at room temperature has been plotted in Fig. 3(d). In the same figure, we plot experimental density also. It is evident from this figure that the density is nearly constant for compositions with x>0.515. Thus, the features shown by elastic modulus are intrinsic to the system and not due to the density variation of the samples. Elastic modulus shows a dip for x=0.530 corresponding to the maximum in dielectric, electromechanical coupling and piezoelectric constants. The various elastic moduli ($1/s_{ij}$) are known to show minimum for x=0.520 in pure PZT also [1] but these measurements were carried out using some indirect methods. Our results based on pulse-echo method are more accurate and direct way of measuring the elastic modulus. Our results prove that the high piezoelectric response in the vicinity of the MPB in PSZT is due to the softening of elastic modulus. This may also be the case for PZT. Softening of elastic modulus indicates instability of the lattice as a function of composition on approaching the MPB. A minima in the elastic modulus at the MPB implies that a small electric field can produce large piezoelectric strain as a result of reduced interatomic force constant for atomic displacements. Softening of elastic modulus is also found to be responsible for high values of piezoelectric response in wurtzite alloys such as $Sc_xAl_{1-x}N$, $Y_xIn_{1-x}N$ etc [45, 46]. Thus, the maximum piezoelectric response in PSZT occurring at x=0.530 may be linked with the elastic instability as a function of composition at room temperature.

It is interesting to note that way back in 1973, using Raman scattering, Pinckzuk [47] reported softening of the transverse optic mode frequency ($\omega_{TO}$) on approaching the MPB composition in pure PZT at room temperature from the tetragonal side. In fact, they found the$\omega_{TO}^2=k(x-x_c)$ type dependence of the soft mode frequency, where $x_c$ is the MPB composition. We believe that the minima in the elastic modulus shown in Fig.3 for PSZT at $x_c$=0.530 is linked with the composition induced softening of the polar optical phonon at q=0 and is responsible for the peak in dielectric constant as per Lydanne-Sachs-Teller relationship [48] where the temperature is replaced by the composition. It is well known that in ferroelectric perovskites and several other materials [49, 50], the primary order parameter polarization is coupled with the secondary order parameter associated with the spontaneous deformation of the lattice (i.e. strain $\eta$). For the cubic to tetragonal ferroelectric transition in perovskites, the secondary order parameter strain has been shown to exhibit quadratic electrostrictive coupling with primary order parameter polarization (P) as $\eta = Q\ P^2$, where Q is the electrostrictive coefficient [51]. This implies the strong softening of the polar phonon mode will lead to a similar softening in the acoustic mode as reflected through our elastic modulus measurements. Our results thus suggest the lattice instability as a function of composition on the approaching the MPB composition at which physical properties show a maximum responses responsible for the enhancement of the piezoelectric response. Thus, the softening of elastic modulus in PSZT and PZT may be the key factor for the maximum piezoelectric response at x=0.530 and 0.520, respectively. Since 6% Sr substitution does not change the topology of the phase diagram including structure of crystallographic phases involved or the peaking of the properties at the MPB, except for shifting the MPB composition at room temperature from x=0.520 in PZT to x=0.530 in PSZT, we believe that our analysis is valid for the family of PZT ceramics in general.

**IV. Conclusions**

In this work, using Rietveld analysis of SXRD data we establish the structure of MPB compositions of PSZT. It is found that the MPB region (Δx) in PSZT is quite narrow and corresponds to $0.520 \leq x \leq 0.535$ with $\Delta x \simeq 0.015$. Further, the MPB region separates the stability fields of ferroelectric tetragonal phase for $x \leq 0.515$ and ferroelectric pseudorhombohedral monoclinic phase in Cm space group for $x \geq 0.545$ and not the stability fields of tetragonal and rhombohedral phases [1]. Our physical property measurements show that $d_{33}$, $k_p$ and $\varepsilon_r$ peak at a morphotropic composition for x=0.530 of PSZT at which the elastic modulus shows a minimum.

Our results show that the models based on phase coexistence, polarization rotation, flat energy surface due to tricritical or second order phase transition and elastic softening associated with the tetragonal to monoclinic phase transition near room temperature may enhance the physical properties in the MPB region in general but cannot explain the peaking of the physical properties at a specific composition (x=0.520 for PZT and x=0.530 for PSZT) in the MPB region. Temperature dependence of longitudinal elastic modulus confirms that $T_{AFD}$ lies nearer to room temperature for $x \simeq 0.573$ and therefore proximity of $T_{AFD}$ to room temperature also does not play a specific role in the maximization of the piezoelectric response at the MPB at x=0.530 in PSZT. Also, in our previous work [6, 19], we show that in PSZT, a tricritical point exists at high temperatures for $x \approx 0.550$. This tricritical point leads to the flatter energy surface at room temperature too but for $x \approx 0.550$, whereas the peak in properties occurs at x=0.530. The observation of a minima in the longitudinal elastic modulus occurring at x=0.530 suggests that the softening of elastic modulus as a function of composition on account of a morphotropic phase transition from tetragonal to a pseudorhombohedral monoclinic phase is responsible for the high value of piezoelectric response at room temperature in the vicinity of the MPB.

**Acknowledgements**

D. Pandey acknowledges financial support from Science and Engineering Research Board (SERB) of India through the award of J. C. Bose National Fellowship grant. D. Pandey and Y. Kuroiwa acknowledge financial support from the Department of Science and Technology (DST), Government of India and Japan Society for the Promotion of Science (JSPS) of Japan under the Indo-Japan Science Collaboration Program. The synchrotron radiation experiments were performed at the BL02B2 beam line of Spring-8 with the approval of Japan Synchrotron Radiation Research Institute (Proposal Nos. 2011A1324 and 2011A0084). T. Suzuki and I. Ishii thank the support (the Grant-in-Aid for Scientific Research on Innovative Areas "Heavy Electrons"(No.20102005)) from the Ministry of Education, Culture, Sports, Science, and Technology of Japan. R.S.S. is thankful to the Department of Science and Technology (DST), India for DST INSPIRE Faculty award (DST/INSPIRE/04/2015/002300).

## Figure Captions

**Fig. 1** Synchrotron powder XRD profiles of the $(111)_{pc}$, $(200)_{pc}$ and $(220)_{pc}$ peaks of PSZT for **(a)** $x$ = 0.515, **(b)** $x$ = 0.520, **(c)** $x$ = 0.525, **(d)** $x$ = 0.530, **(e)** $x$ = 0.535, **(f)** $x$ = 0.545, and **(g)** $x$ = 0.550.

**Fig.2** Compositional dependence of **(a)** dielectric constant ( $\varepsilon'$ ), **(b)** planar electromechanical coupling coefficient ($k_p$), **(c)** piezoelectric strain coefficient ($d_{33}$) of PSZT [present work] and PZT [ref. 29] samples, **(d)** longitudinal elastic modulus ($C_L$) and density (ρ) of PSZT samples, **(e)** variation of antiferrodistortive phase transition temperature ($T_{AFD}$) with composition (x) for PSZT obtained from the temperature dependence of $C_L$ [Data point for x=0.530 and 0.550 were taken from refs. 19] and PZT (data points were taken from ref. 5 and 29).

**Fig. 3** Variation of **(a)** the phase fraction with composition (x) for compositions close to the MPB of PSZT. Filled square, filled circle and open circle represent the phase fractions of the tetragonal, pseudotetragonal monoclinic and pseuodthombohedral monoclinic phases; **(b)** unit cell parameters for phase 1 and phase 2 and **(c)** $c_{p1}/a_{p1}$ ratio and volume ($V_{p1}$) of phase 1 and $c_{p2}/a_{p2}$ ratio and volume ($V_{p2}$) of phase 2 with composition (x) for PSZT. Phase 1 is the pseudorhombohedral monoclinic phase. Phase 2 is tetragonal for x≤0.525 and pseudotetragonal monoclinic for x>0.525.

**Fig. 4** Temperature dependence of longitudinal elastic modulus ($C_L$) for **(a)** x=0.515, **(b)** 0.520, **(c)** 0.525 and **(d)** 0.545.

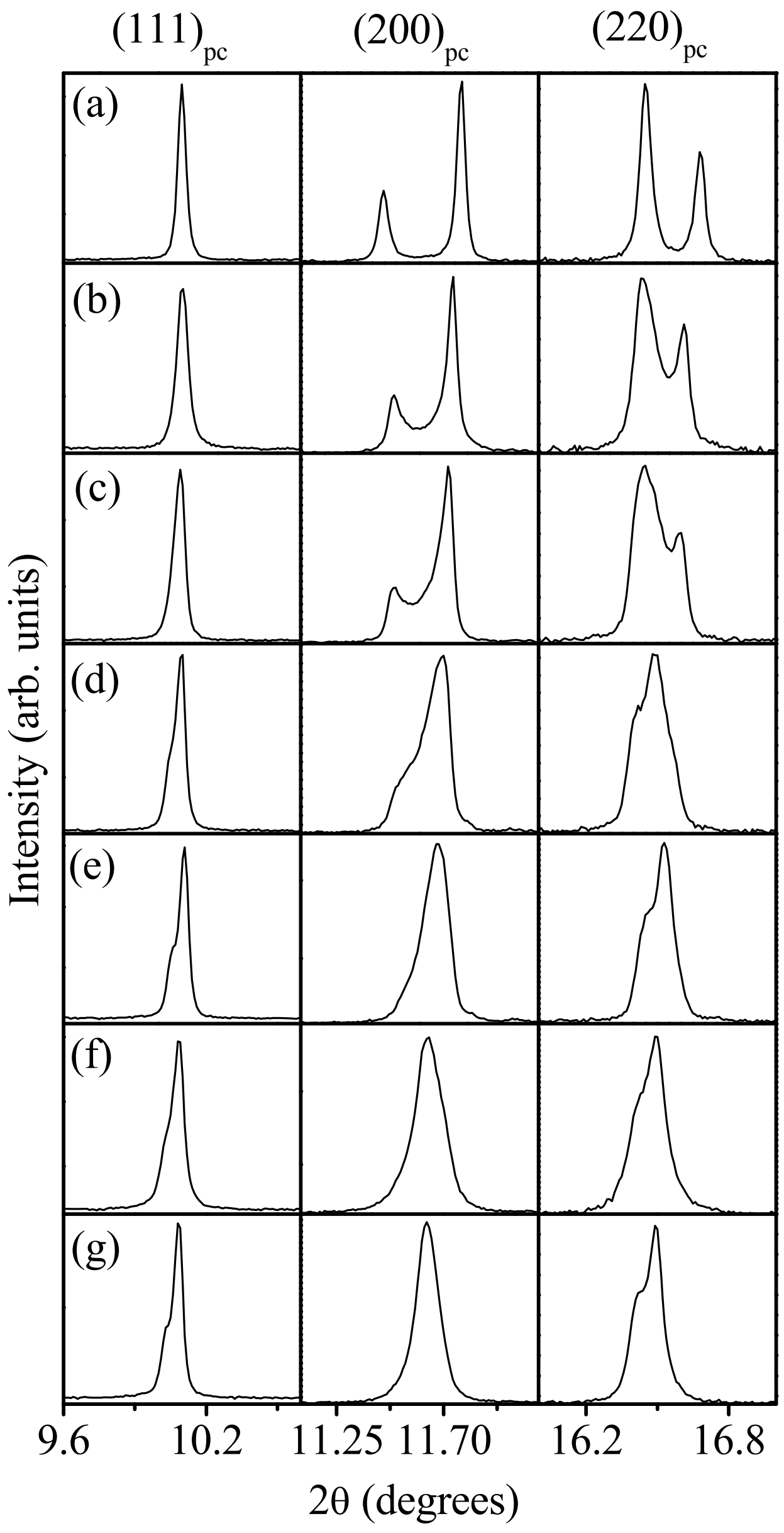

**Fig. 1**

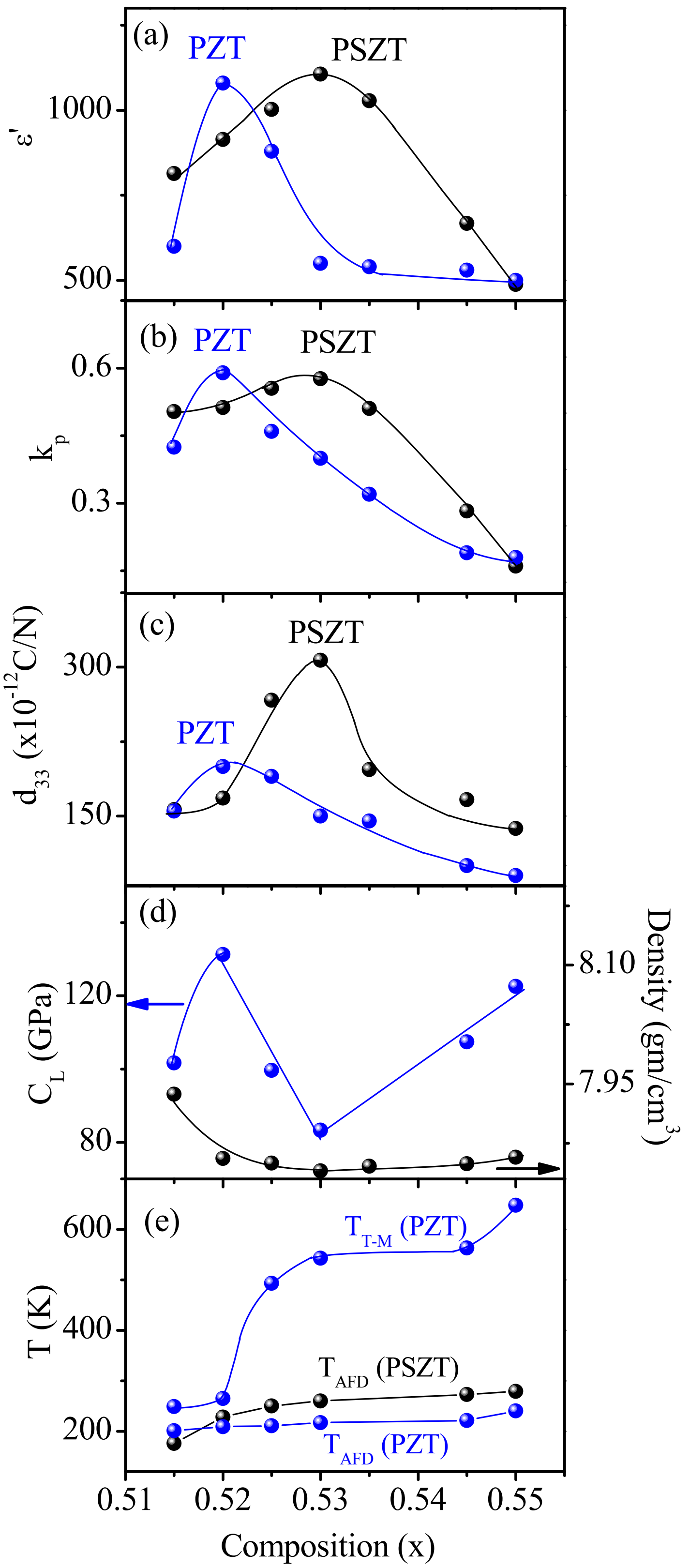

**Fig. 2**

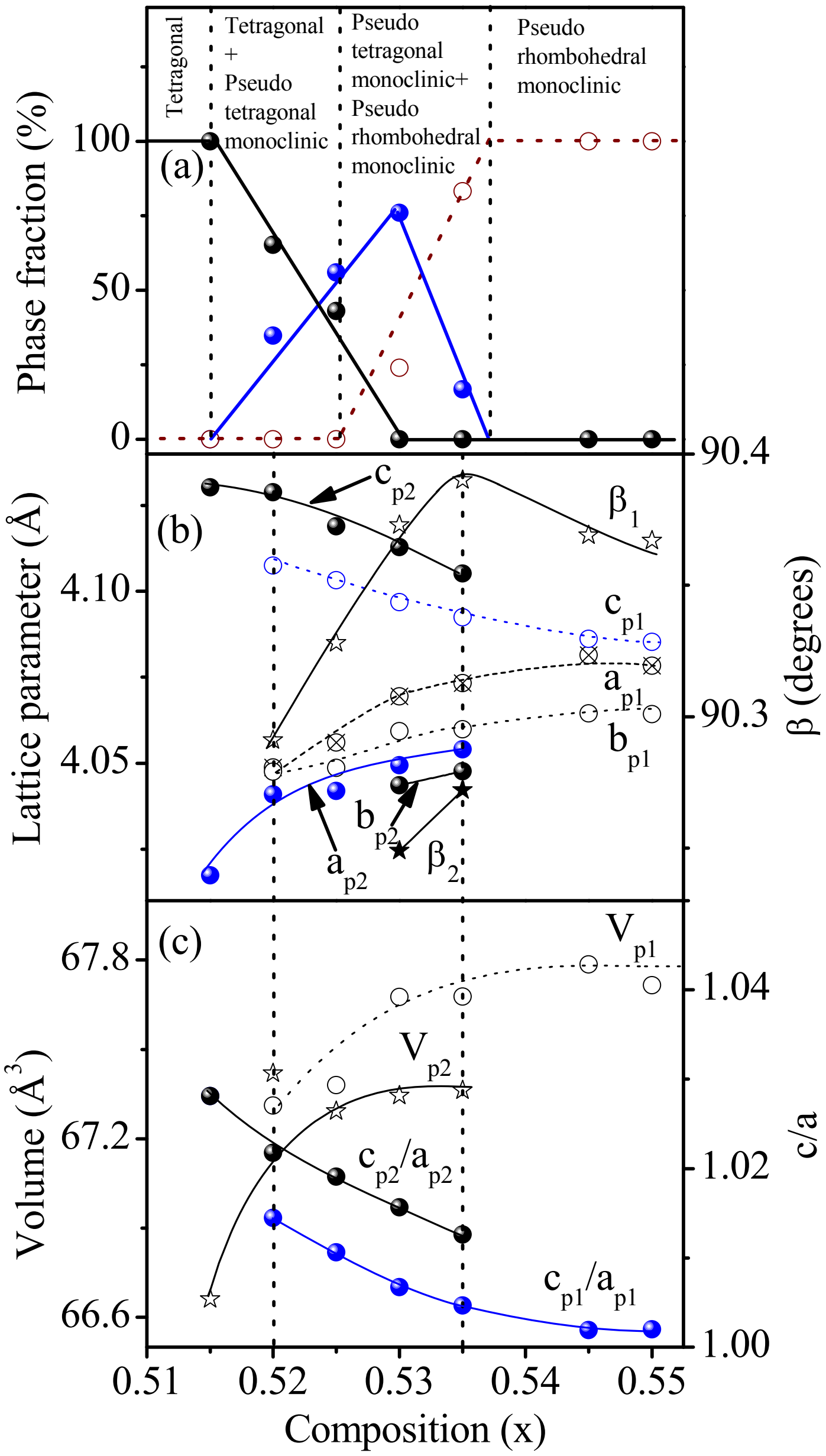

Tetragonal
Tetragonal + Pseudo tetragonal monoclinic
Pseudo tetragonal monoclinic+ Pseudo rhombohedral monoclinic
Pseudo rhombohedral monoclinic
Phase fraction (%)
(a)
100
50
0
(b)
Lattice parameter (Å)
4.10
4.05
β (degrees)
90.4
90.3
c p2
β 1
c p1
a p1
b p1
b p2
a p2
β 2
(c)
Volume (Å^3)
67.8
67.2
66.6
c/a
1.04
1.02
1.00
V p1
V p2
c p2/a p2
c p1/a p1
0.51
0.52
0.53
0.54
0.55
Composition (x)

**Fig. 3**

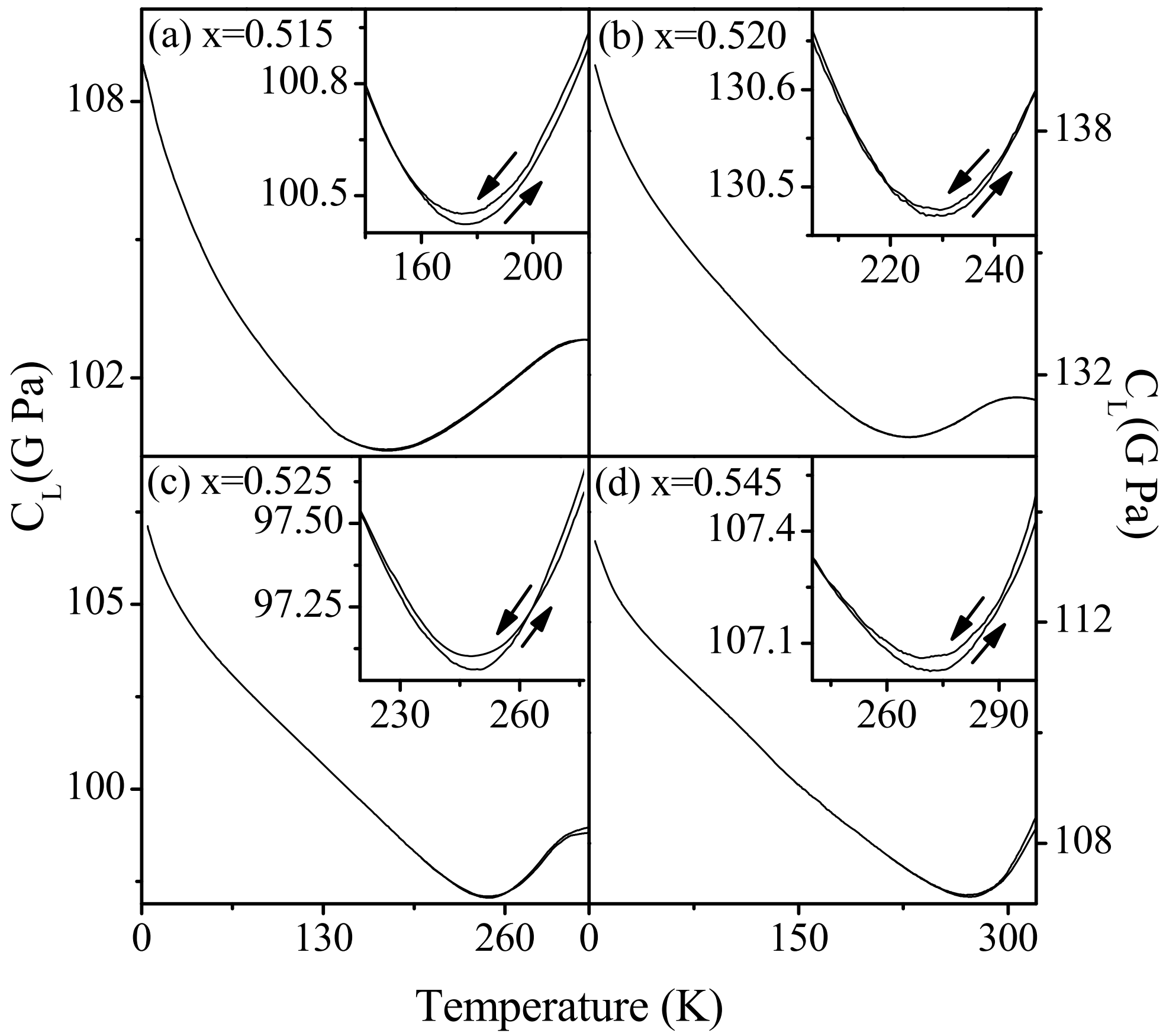

**Fig. 4**

## Supplementary information

**Origin of high piezoelectricity at the morphotropic phase boundary (MPB) in $(Pb_{0.940}Sr_{0.06})(Zr_xTi_{1-x})O_3$**

Ravindra Singh Solanki,[1] Sunil Kumar Mishra,[2] Chikako Moriyoshi,[3] Yoshihiro Kuroiwa,[3] Isao Ishii,[4] Takashi Suzuki,[4] and Dhananjai Pandey[1]

[1]School of Materials Science and Technology, Indian Institute of Technology (Banaras Hindu University), Varanasi-221005, India

[2]UGC-Academic Staff College, Dr. H. S. Gour Vishwavidyalaya, Sagar-470003, Madhya Pradesh, India

[3]Department of Physical Science, Graduate School of Science, Hiroshima University, Japan

[4]Department of Quantum Matter, ADSM, Hiroshima University, Japan

## S.1: Microstructure

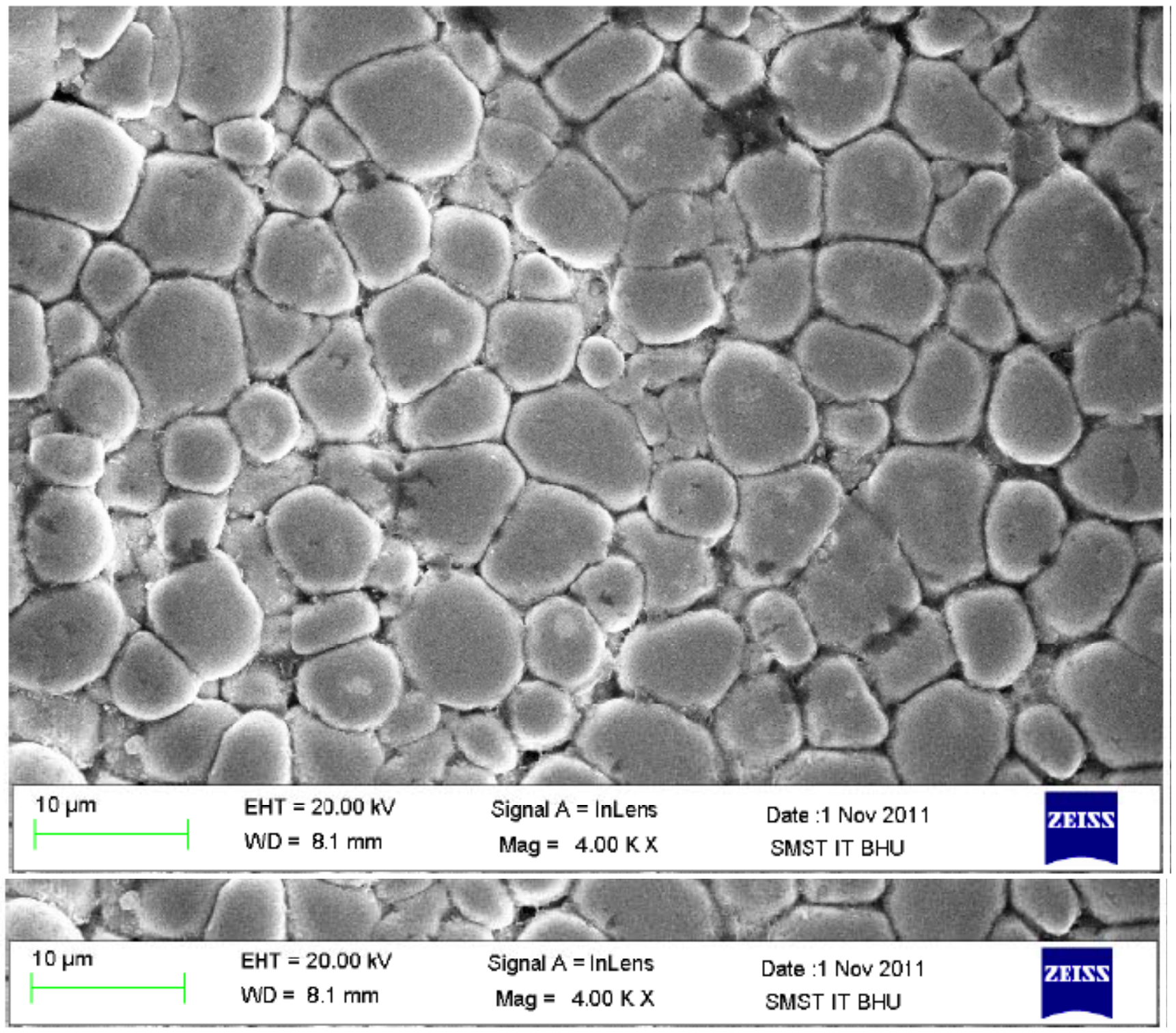

**Fig. S1:** Scanning electron micrographs (SEM) of PSZT ceramic for x = 0.530. For SEM analysis sintered pellets were polished and chemically etched using 2% HF solution in HCI. Thereafter, sintered pellets were coated with conducting gold film by sputtering under vacuum before recording the SEM images.

## S.2: Details of Rietveld refinement

The Rietveld refinement of the tetragonal $(Pb_{0.94}Sr_{0.06})(Zr_{0.515}Ti_{0.485})O_3$ (PSZT515) phase was carried out using P4mm space group with isotropic as well as anisotropic thermal parameters for $Pb^{2+}$ following Noheda et al. [1] and Ragini et al. [2]. Use of isotropic thermal parameters led to reasonable Rietveld fit but the thermal parameter ($B_{iso}$=1.66 Å$^2$) corresponds to a displacement of ~0.145Å for $Pb^{2+}$ which is quite large in comparison to other atoms. Similar high values of $B_{iso}$ for Pb reported in the literature in other Pb-based perovskite oxides including pure PZT have been ascribed to random local displacement of $Pb^{2+}$ cation [see e.g.

ref. 1-3]. For pure PZT, Noheda et al. [1] and Ragini et al. [2] have shown that the off-centre displacement of $Pb^{2+}$is in <110> directions. A similar situation exists in PSZT. The correct value of the local displacement in the $<110>_{pc}$ direction leads to a minima in the $\chi^2$ versus local displacement plot (see Fig S.2.1(b) of supplementary file). Rietveld fits with local atomic disorder are shown in Fig. S.2.1(a) of the supplementary file and the insets of this figure shows the magnified view of the observed and calculated profiles of the $(111)_{pc}$, $(200)_{pc}$ and $(220)_{pc}$ perovskite reflections. The fits are evidently quite good. Table I of the supplementary file gives the refined structural parameters and agreement factors for PSZT515. The presence of local $Pb^{2+}$ disorder in the $<110>_{pc}$ direction clearly suggests the presence of local monoclinic distortion that gives rise to pseudotetragonal monoclinic phase on increasing the $Zr^{2+}$ content further [1].

Based on the singlet like character of the $(111)_{pc}$ peak and doublet like nature of $(200)_{pc}$ and $(220)_{pc}$ peaks, Rietveld refinement was carried out using tetragonal P4mm space group, like that for x=0.515, for x=0.520 also but the fit between the observed and calculated profile is rather unsatisfactory with $\chi^2$=3.55 ( see Fig. S.2.2(a) of supplementary file). Since there is a marked increase in the FWHM of the $(111)_{pc}$for x=0.520 as compared to that for x=0.515, the $(111)_{pc}$ may not be a singlet. In that situation, the monoclinic Cm space group becomes plausible, as, for the Cm space group, none of the $(h00)_{pc}$, $(hh0)_{pc}$ and $(hhh)_{pc}$ peaks is a singlet. Accordingly, Rietveld refinement was carried out using Cm space group also. While the fit between the observed and calculated profiles has improved a little with lower $\chi^2$ value (3.27) as compared to that (3.55) for the P4mm space group, the $(h00)_{pc}$ and $(hh0)_{pc}$ peaks (see the inset of Fig. S.2.2(b) of supplementary file) show poor fits. As a next step, we considered coexistence of tetragonal (P4mm) and monoclinic (Cm) phases in the Rietveld refinements. This led to a very significant improvements in the fits as shown in Fig. S.2.2(c) of supplementary file and its inset with a much lower value of $\chi^2$ (1.17). We thus conclude that both the P4mm and Cm phases coexist for this composition. Similarly, the refinements for PSZT525 were carried out using P4mm+Cm structural model. Table I of the supplementary file lists the refined structural parameters and agreement factors for PSZT520 and PSZT525 using P4mm+Cm phase coexistence model. The equivalent perovskite cell parameters $a_p \approx a_m/\sqrt{2}$=4.04880Å, $b_p \approx b_m/\sqrt{2}$=4.0475Å and $c_p = c_m$=4.1075Å of the Cm phase for x=0.520 bear pseudotetragonal relationship ($a_p \approx b_p \neq c_p$). A similar pseudotetragonal relationship is observed for x=0.525 also. This pseudotetragonality is similar to pure PZT with x=0.525 [4].

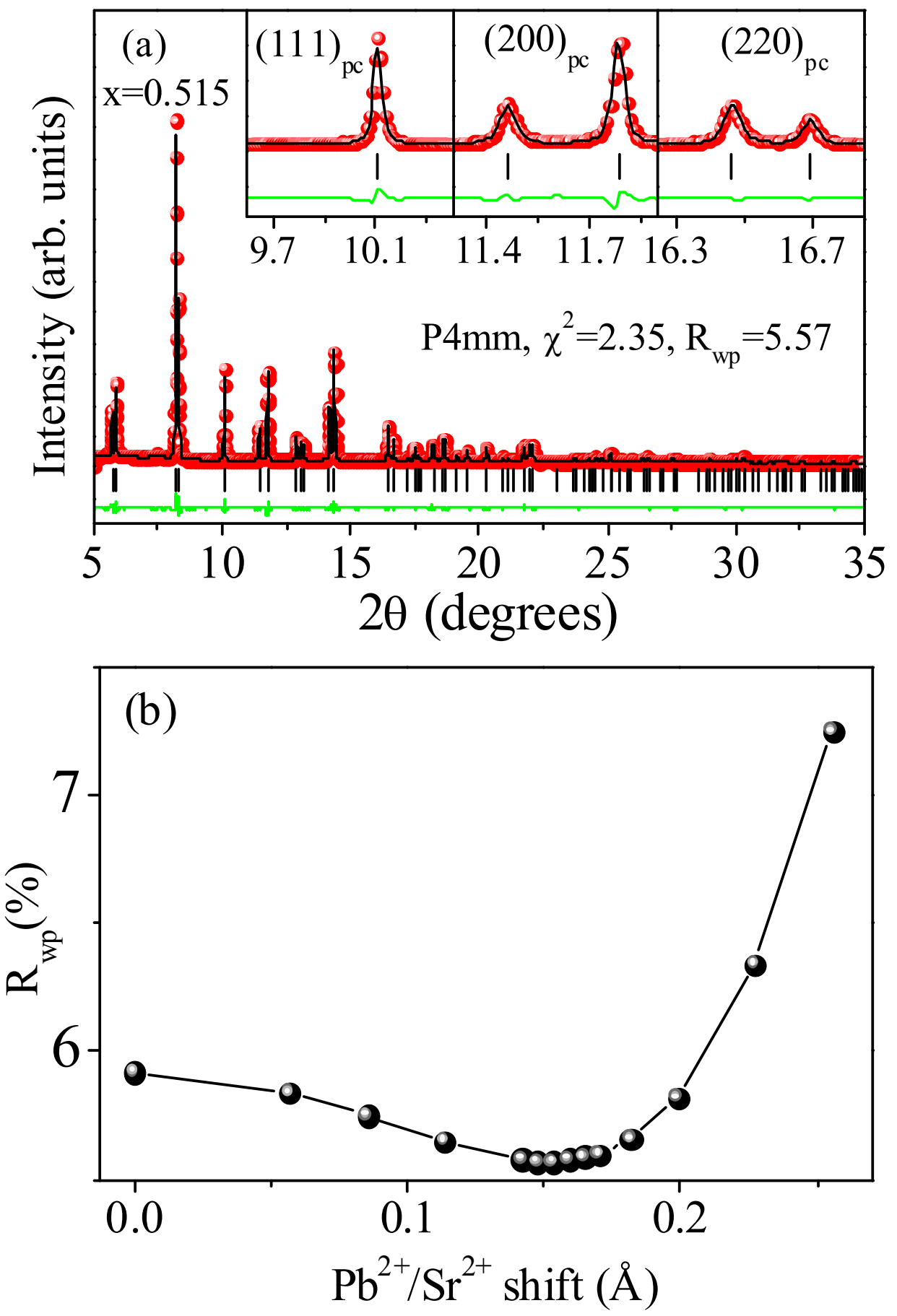

**Fig. S.2.1 (a)** Observed (red filled circles), calculated (black continuous line), and difference (bottom line) patterns at room temperature for x= 0.515 using P4mm space group with local disorder for $Pb^{2+}/Sr^{2+}$. The vertical tick marks above the difference line stand for the Bragg peak positions and (b) Variation of the agreement factor $R_{wp}$ as a function of $Pb^{2+}/Sr^{2+}$ shifts for refinements of PSZT515 with various fixed values of displacements along tetragonal <110> direction.

It is because of the pseudotetragonality of the structure, this phase is termed as pseudotetragonal monoclinic following the nomenclature on PZT [1, 4, 5]. We believe that the coexistence of the monoclinic phase in the Cm space group with the tetragonal phase in P4mm space group forces the former to adopt pseudotetragonal cell parameters ($a_p \approx b_p < c_p$) to minimize the strain energy at the inter-phase interface [6]. Phase fractions of the pseudotetragonal monoclinic phase increases from ~35±1 % for x=0.520 to 56±1 %for x=0.525 with a corresponding decrease in the phase fraction of the tetragonal phase.

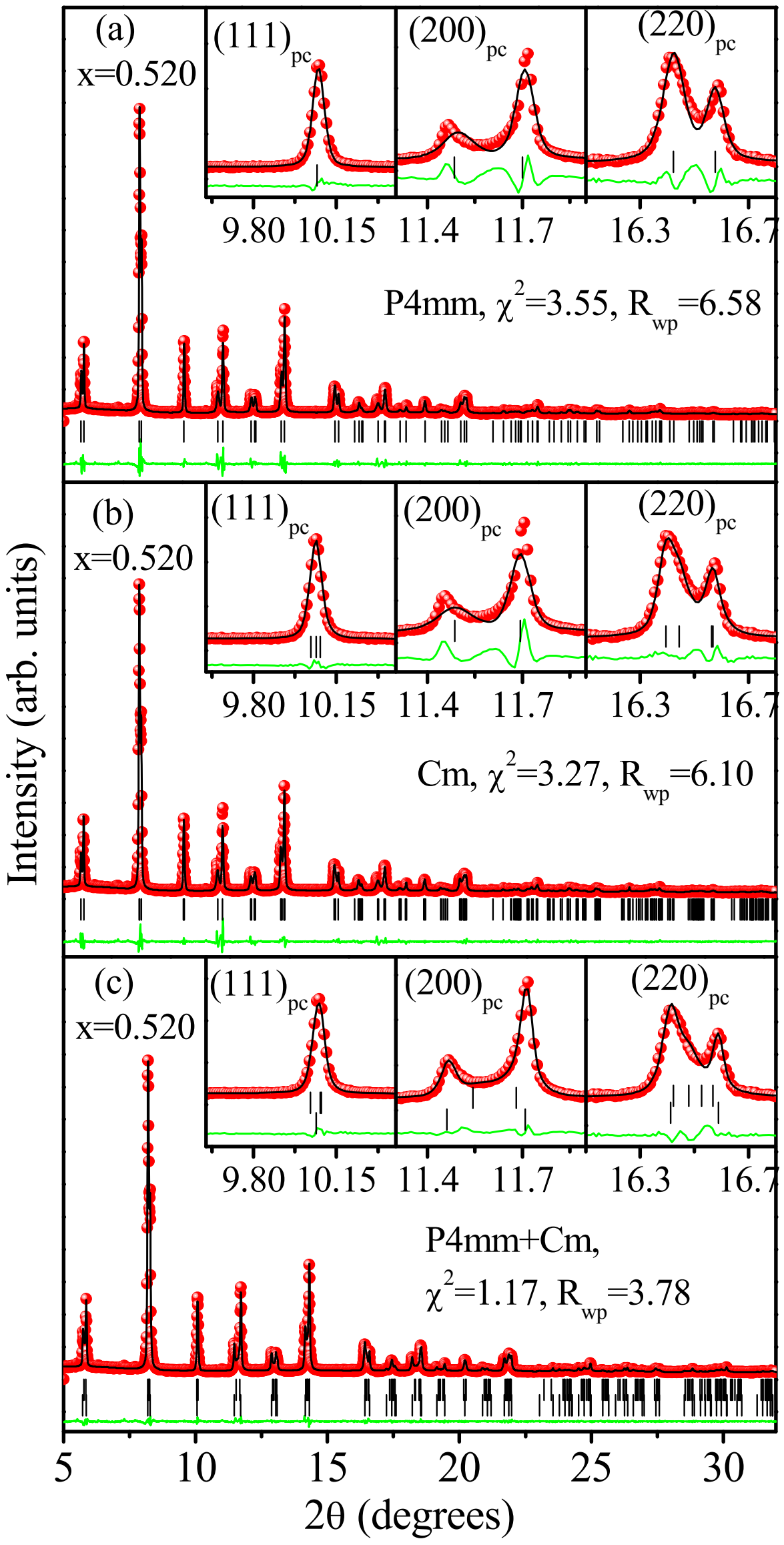

**Fig. S.2.2** Observed (red filled circles), calculated (solid line), and difference (bottom line) patterns obtained from the Rietveld analysis of the room temperature synchrotron x-ray diffraction data of PSZT520 using (a) P4mm, (b) Cm and (c) P4mm+Cm structural models. The vertical tick marks above the difference line stand for the Bragg peak positions.

Variation of the phase fraction with composition (x) has been plotted in Fig. 3 (a) of main text. The crystal structure of PZT for x≥0.545 with singlet like $(200)_{pc}$ peak and doublet $(111)_{pc}$ peak has been historically regarded as rhombohedral [7] in the R3m space group. However, as mentioned earlier, Ragini et al. [2] and Singh et al. [5] showed that the FWHM of $(200)_{pc}$ peak is much larger than the FWHM of nearby $(111)_{pc}$ reflection for x≥0.530 that cannot be accounted for using anisotropic strain broadening model of Stephen's [8].

Excellent fit between observed and calculated profiles was observed for such compositions using a pseudorhombohedral monoclinic phase in Cm space group at room temperature [2, 5]. We have discussed this aspect in our previous comprehensive study of a pseudorhombohedral monoclinic composition of PSZT for x=0.550 [9]. Rietveld refinements for PSZT with x=0.545 using single phase Cm space group model also give the same results. Table I of the supplementary file lists the refined structural parameters and agreement factors for PSZT with x=0.545 and 0.550 along with other compositions of PSZT. The equivalent perovskite cell parameters $a_p \approx a_m/\sqrt{2}$, $b_p \approx b_m/\sqrt{2}$ and $c_p \approx c_m$ of the Cm phase reveal $a_p \approx b_p \approx c_p$ for x=0.545. A similar situation holds good for x=0.550 also. This is the characteristic of the pseudorhombohedral monoclinic phase discussed in ref. 6. So the structure of PSZT for x≥0.545 is pseudorhomohedral monoclinic. We now return to the composition range 0.525<x<0.545 as it is also found to be a two phase region. The linear extrapolation of phase fraction of tetragonal phase for x=0.520 and 0.525 suggests that this phase should be absent in PSZT530 (See Fig. 3 (a) of main text). Accordingly, we carried out refinements for PSZT530 using single Cm phase like for x ≥ 0.545. Rietveld fits are shown in Fig. S.2.3 (a) of supplementary file for this structural model, however, reveal that the peak positions (see e.g. $(200)_{pc}$ peaks in the inset) are not correctly accounted for. We therefore applied the P4mm+Cm phase coexistence model of x≤0.525 compositions in the Rietveld refinements for PSZT530 also. The fits corresponding to this model are shown in Fig. S.2.3(c) of supplementary file which reveals a better fit as compared to single Cm phase model. But the P4mm+Cm coexistence model gives higher phase fraction (62%) of the tetragonal phase as compared to that for x=0.525. This is physically unrealistic as the phase fraction of the tetragonal phase is expected to decrease until the structure becomes pure monoclinic for x≥0.545 as can be seen from Fig. 3(a) of main text. We can thus rule out the P4mm+Cm phase coexistence model at room temperature for x=0.530. A similar situation holds good for PSZT535 also. We then considered a model based on the coexistence of pseudotetragonal and pseudorhombohedral monoclinic phases and this improved the fits and lowered the $\chi^2$ value from 1.66 for the pseudotetragonal Cm+ pseudorhombohedral Cm structural model to 1.15. We thus conclude that PSZT530 contains coexisting pseudotetragonal and pseudorhombohedral monoclinic phases, both in the Cm space group (Fig. S.2.3(b)). Table I of the supplementary file lists the refined structural parameters and agreement factors for PSZT530 and 535. The equivalent perovskite cell parameters $a_p \approx a_m/\sqrt{2}$ =4.05191Å, $b_p \approx b_m/\sqrt{2}$ =4.03948Å and $c_p = c_m$=4.1149Å show pseudotetragonal relationship ($a_p \approx b_p \neq c_p$) while $a_p \approx a_m/\sqrt{2}$=4.06945Å, $b_p \approx b_m/\sqrt{2}$=4.05933Å and $c_p$=$c_m$=4.09684Å of the second Cm phase reveals pseudorhombohedral relationship ($a_p \approx b_p \approx c_p$) for PSZT530. Lattice parameters of PSZT535 also show similar relationships. As shown in Fig 3(a) of the main text, the phase fraction of the pseudorhombohedral monoclinic phase increases on increasing $Zr^{4+}$ content from x=0.530 to 0.535 while that of the pseudotetrgonal phase decreases.

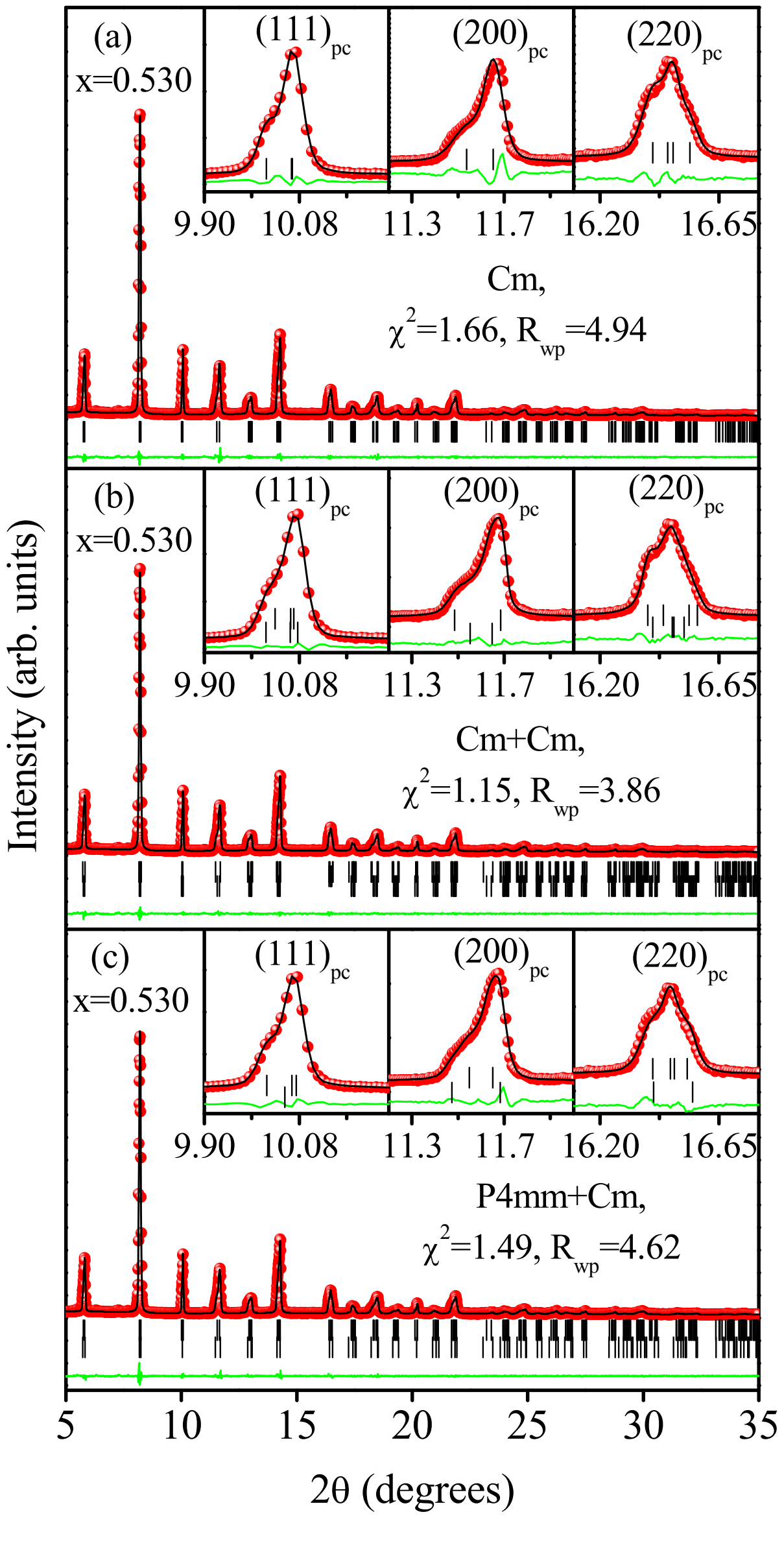

**Fig. S.2.3** Observed (red filled circles), calculated (solid line), and difference (bottom line) patterns obtained from the Rietveld analysis of the room temperature synchrotron x-ray diffraction data of PSZT530 using (a) Cm, (b) Cm+Cm and (c) P4mm+Cm structural models. The vertical tick marks above the difference line stand for the Bragg peak positions.

**Table I:** Refined structural parameters of $(Pb_{0.94}Sr_{0.06})(Zr_xTi_{1-x})O_3$ for $0.515 \leq x \leq 0.550$

| x | 0.515 | | 0.520 | | 0.525 | | 0.530 | | 0.535 | | 0.545 | 0.550 |
|---|---|---|---|---|---|---|---|---|---|---|---|---|
| Space group | P4mm | | Two phase compositions | | | | | | | | Cm (pseudorh-ombohedral) | Cm (pseudorh-ombohedral) |
| | Anisotropic thermal parameter model | Local shift of $Pb^{2+}$ along $<110>_{pc}$ | P4mm | Cm (pseudotet-ragonal ) | P4mm | Cm (pseudotet-ragonal) | Cm (pseudotet-ragonal) | Cm (pseudorh-ombohedral) | Cm (pseudotet-ragonal) | Cm (pseudorh-ombohedral) | | |
| a(Å) | 4.0174 (2) | 4.0174 (2) | 4.0409(8) | 5.7250(4) | 4.0409(5) | 5.7352(4) | 5.7294(4) | 5.7542(3) | 5.7300(1) | 5.7487(3) | 5.7711(3) | 5.7668(3) |
| b(Å) | 4.0174 (2) | 4.0174 (2) | 4.0409(8) | 5.7217(5) | 4.0409(5) | 5.7248(3) | 5.7182(4) | 5.7399(3) | 5.7233(9) | 5.7350(2) | 5.7472(2) | 5.7468(2) |
| c(Å) | 4.13030(2) | 4.13030(2) | 4.12889(1) | 4.1075(4) | 4.1190(1) | 4.1022(2) | 4.1149(3) | 4.09684(2) | 4.1052(7) | 4.0865(2) | 4.0893(2) | 4.0853(2) |
| β (degrees) | 90 | 90 | 90.00 | 90.291(6), | 90.00 | 90.318(4) | 90.249(6) | 90.373(2) | 90.27(7) | 90.398(2) | 90.369(2) | 90.367(2) |
| Pb/Sr (x=y=z) | 0.00 | x=y=0.026, z=0.00 | 0.00 | 0.00 | 0.00 | 0.00 | 0.00 | 0.00 | 0.00 | 0.00 | 0.00 | 0.00 |
| Zr/Ti (x) | 0.50 | 0.50 | 0.00 | 0.513(3) | 0.00 | 0.528(1) | 0.507(5) | 0.511(2) | 0.49(1) | 0.516(2) | 0.516(1) | 0.520(1) |
| Zr/Ti (z) | 0.5495(7) | 0.5495(7) | 0.5515(7) | 0.460(2) | 0.5446(8) | 0.459(2) | 0.453(4) | 0.453(2) | 0.433(3) | 0.461(2) | 0.458(1) | 0.458(1) |
| O1(x) | 0.50 | 0.50 | 0.50 | 0.565(4) | 0.50 | 0.570(4) | 0.54(1) | 0.534(7) | 0.60(2) | 0.547(5) | 0.539(5) | 0.518(6) |
| O1(z) | 0.091(4) | 0.091(4) | 0.092(3) | -0.084(3) | 0.099(4) | -0.075(3) | -0.11(1) | -0.058(7) | -0.14(2) | -0.071(5) | -0.064(5) | -0.059(5) |
| O2(x) | 0.50 | 0.50 | 0.50 | 0.29(1) | 0.50 | 0.30(1) | 0.299(9) | 0.302(4) | 0.31(2) | 0.284(3) | 0.277(3) | 0.286(3) |
| O2(y) | 0.50 | 0.50 | 0.00 | 0.28(1) | 0.00 | 0.274(9) | 0.207(8) | 0.259(4) | 0.32(1) | 0.234(3) | 0.245(4) | 0.252(3) |
| O2(z) | 0.602(2) | 0.602(2) | 0.595(2) | 0.48(2) | 0.584(3) | 0.46(1) | 0.433(1) | 0.405(4) | 0.43(1) | 0.418(3) | 0.388(2) | 0.394(3) |
| B (Pb/Sr ) | $\beta_{11}$=0.0327(5) $\beta_{22}$=0.0327(5) $\beta_{33}$=0.0157(7) | 1.019 | $\beta_{11}$=0.0351(5) $\beta_{22}$=0.0351(5) $\beta_{33}$=0.0135(7) | $\beta_{11}$=0.031(2) $\beta_{22}$=0.018(2) $\beta_{33}$=0.028(2) $\beta_{13}$=0.016(2) | $\beta_{11}$=0.035(6) $\beta_{22}$=0.035(6) $\beta_{33}$=0.016(8) | $\beta_{11}$=0.038(2) $\beta_{22}$=0.019(1) $\beta_{33}$=0.033(2) $\beta_{13}$=0.017(1) | $\beta_{11}$=0.025(3) $\beta_{22}$0.0290(3) $\beta_{33}$=0.088(5) $\beta_{13}$=0.005(4) | $\beta_{11}$=0.015(1) $\beta_{22}$=0.016(2) $\beta_{33}$=0.005(1) $\beta_{13}$=0.012(8) | $\beta_{11}$=0.016(6) $\beta_{22}$=0.029(5) $\beta_{33}$=0.026(5) $\beta_{13}$=0.019(6) | $\beta_{11}$=0.023(1) $\beta_{22}$=0.016(9) $\beta_{33}$=0.023(1) $\beta_{13}$=0.013(9) | $\beta_{11}$=0.018(1) $\beta_{22}$=.0124(2) $\beta_{33}$=0.022(1) $\beta_{13}$=.0128(4) | $\beta_{11}$=0.0150(1) $\beta_{22}$=0.0180(1) $\beta_{33}$=0.0170(2) $\beta_{13}$=0.0158(2) |
| B (Zr/Ti) | 0.002 | 0.002 | 0.007 | 0.007 | 0.011 | 0.011 | 0.014 | 0.014 | 0.04 | 0.04 | 0.002 | 0.002 |
| B (O1) | 0.545(3) | 0.545(3) | 0.55(3) | 0.02(2) | 0.90(2) | 0.60(1) | 1.55(2) | 1.02(3) | 0.04(2) | 0.02(3) | 1.45(3) | 1.55(2) |
| B (O2) | 0.718(1) | 0.718(1) | 1.06(5) | 0.10(1) | 0.53(4) | 0.50(5) | 0.09(2) | 0.07(1) | 0.014(2) | 0.07(1) | 0.031(1) | 0.09(2) |
| Phase fraction | 100 | 100 | 65.22 | 34.78 | 43.80 | 56.20 | 75.88 | 24.12 | 16.73 | 83.27 | 100 | 100 |
| $R_{wp}$ | 5.56 | 5.57 | 3.78 | | 3.52 | | 4.52 | | 4.86 | | 2.73 | 2.73 |
| $\chi^2$ | 2.35 | 2.35 | 1.17 | | 1.32 | | 1.54 | | 1.60 | | 2.92 | 2.95 |